\documentclass[aps,twocolumn,superscriptaddress,reprint]{revtex4-1}

\usepackage{times}
\usepackage{xfrac}

\usepackage{amsmath}
\usepackage{amsfonts}
\usepackage{amssymb}
\usepackage[caption=false]{subfig}
\usepackage{graphicx}
\usepackage{color}
\usepackage{float}
\usepackage[T1]{fontenc}

\def\<{\langle}
\def\>{\rangle}
\newcommand\opone{\leavevmode\hbox{\small1\kern-3.8pt\normalsize1}}
\newcommand{\e}{{\rm e}}

\AtBeginDocument{}
\usepackage{url}
\usepackage[unicode=true, breaklinks=false, pdfborder={0 0 1}, backref=false, colorlinks=true, linkcolor=blue, citecolor=blue]{hyperref}
\expandafter\def\expandafter\UrlBreaks\expandafter{\UrlBreaks
  \do\a\do\b\do\c\do\d\do\e\do\f\do\g\do\h\do\i\do\j%
  \do\k\do\l\do\m\do\n\do\o\do\p\do\q\do\r\do\s\do\t%
  \do\u\do\v\do\w\do\x\do\y\do\z\do\A\do\B\do\C\do\D%
  \do\E\do\F\do\G\do\H\do\I\do\J\do\K\do\L\do\M\do\N%
  \do\O\do\P\do\Q\do\R\do\S\do\T\do\U\do\V\do\W\do\X%
  \do\Y\do\Z\do\1\do\2\do\3\do\4\do\5\do\6\do\7\do\8%
  \do\9\do\0}
  \usepackage{subfig}

\usepackage{xcolor}
\usepackage[normalem]{ulem}
\newcommand{\stkout}[1]{\ifmmode\text{\sout{\ensuremath{#1}}}\else\sout{#1}\fi}

\begin{document}

\title{Entropic bounds on information backflow}

\date{\today}

\author{Nina Megier}
\email{nina.megier@mi.infn.it}
\affiliation{Dipartimento di Fisica ``Aldo Pontremoli'', Universit\`a degli Studi di Milano, via Celoria 16, 20133 Milan, Italy}
\affiliation{Istituto Nazionale di Fisica Nucleare, Sezione di Milano, via Celoria 16, 20133 Milan, Italy}
\author{Andrea Smirne}
\affiliation{Dipartimento di Fisica ``Aldo Pontremoli'', Universit\`a degli Studi di Milano, via Celoria 16, 20133 Milan, Italy}
\affiliation{Istituto Nazionale di Fisica Nucleare, Sezione di Milano, via Celoria 16, 20133 Milan, Italy}
\author{Bassano Vacchini}
\affiliation{Dipartimento di Fisica ``Aldo Pontremoli'', Universit\`a degli Studi di Milano, via Celoria 16, 20133 Milan, Italy}
\affiliation{Istituto Nazionale di Fisica Nucleare, Sezione di Milano, via Celoria 16, 20133 Milan, Italy}

\begin{abstract}
  In the dynamics of open quantum systems, the backflow of information
  to the reduced system under study has been suggested as the actual
  physical mechanism inducing memory and thus leading to non-Markovian
  quantum dynamics. To this aim, the trace-distance or Bures-distance
  revivals between distinct evolved system states have been shown to
  be subordinated to the establishment of system-environment
  correlations or changes in the environmental state. We show that
  this interpretation can be substantiated also for a class of
  entropic quantifiers. We exploit a suitably regularized version of
  Umegaki's quantum relative entropy, known as telescopic relative
  entropy, that is tightly connected to the quantum Jensen-Shannon
  divergence. In particular, we derive general upper bounds on the
  telescopic relative entropy revivals conditioned and determined by
  the formation of correlations and changes in the environment. We
  illustrate our findings by means of examples, considering the
  Jaynes–Cummings model and a two-qubit dynamics.
\end{abstract}

\maketitle

\section{Introduction}

The notion of quantum non-Markovianity has been attracting a lot of
interest for more than a decade now \cite{Breuer_2012,Rivas_2014,
  Breuer2016a}. In this time, a wide variety of different definitions
of quantum non-Markovianity have been proposed, all of them aimed to
{reveal} the occurrence of memory effects in quantum evolutions. The
most widespread are the ones based on the divisibility property of the
dynamical map \cite{Wolf2008a,PhysRevLett.105.050403,
  PhysRevA.89.042120, HelstromPdiv}, the monotonicity of the trace
distance (TD) between two distinct reduced states \cite{BLP, Helstrom,
  HelstromPdiv}, the change of the volume of accessible reduced states
\cite{PhysRevA.88.020102}, and the process tensor formalism
\cite{PhysRevA.97.012127, PhysRevLett.123.040401,PhysRevA.100.012120}.
Furthemore, also entropic quantities have been used to detect
non-Markovianity, see \cite{PhysRevLett.112.210402,
  PhysRevA.101.020303}.  The interest toward non-Markovian quantum
dynamics has not only theoretical motivations: non-Markovianity has
proven to be beneficial, among others, in {quantum control}
\cite{bylicka2014, Reich2015} and teleportation tasks
\cite{Laine2014a}.  As the different definitions of quantum
non-Markovianity are in general not equivalent, it is all the more
important to find their corresponding physical interpretations.
In the approach based on TD, its increase in time
signifies a backflow of information to the open system, resulting in
an enhanced reduced-state distinguishability and representing the
distinctive trait of memory effects in the dynamics.  The revivals of
distinguishability are related to the establishment of
system-environment correlations and changes in the environmental state
\cite{Laine2010b,Mazzola2012,Smirne2013,Cialdi2014,Campbell_2019},
which then appear to be the basic elements ruling the non-Markovian
character of open-system dynamics, though the precise assessment of
their role is still under vivid debate
\cite{megier2017,Breuer_2018,PhysRevA.97.052133,DeSantis2019a,DeSantis2020a,DeSantis2020b,banacki2020information}.

The proof of the connection of the distinguishability revivals with
correlations and environmental state changes as formulated via the TD
essentially relies on the triangular inequality, so that it might be
natural to think that it only holds when distance quantifiers are
used.  Here we show that, to the contrary, such a connection can be
{maintained} also when considering entropic distinguishability
quantifiers.  Our analysis extends and strengthens the viewpoint that
quantum non-Markovianity can be understood in terms of a backflow of
information, which induces an increase of the distinguishability among
open-system states and is microscopically motivated by the generation
of correlations and changes in the environmental state due to the
system-environment interaction.  More specifically, we prove an upper
bound for the revivals in time of a whole class of entropic
distinguishability quantifiers, namely the telescopic relative entropy
(TRE), firstly introduced in \cite{Audenaert1} and providing
regularized versions of the quantum relative entropy
(QRE). Remarkably, our bound allows to quantitatively link the
distinguishability revivals with the establishment of correlations
between the system and the environment due to their mutual
interaction, as well as to the modification of the state of the
environment.  We also focus on a special case of the symmetrised
version of TRE, which coincides with the quantum Jensen-Shannon
divergence (QJSD) \cite{PhysRevA.72.052310}. QJSD is a widely used
distinguishability measure \cite{ PhysRevLett.105.040505,
  PhysRevA.84.032120, PhysRevE.88.032806, Domenico2015, BAI2015344,
  PhysRevLett.116.150504, YE2020113690, SLAOUI2020124946}, for which
it was only recently shown that its square root is a proper distance
\cite{virosztek2019metric, sra2019metrics}. We show that the upper
bound to distinguishability revivals in this case significantly
simplifies and becomes tighter.  Finally, we showcase our findings on
two examples, namely the paradigmatic Jaynes–Cummings model and a two
qubit dynamics, which allows us to compare explicitly the behaviors of
the different quantifiers of distinguishability involved in our
analysis.

The rest of the paper is organized as follows. In Sec.\ref{sec:ibm}, we briefly recall 
the main features of the TD characterization of non-Markovianity that are relevant to our analysis.
In Sec.\ref{sec:tre}, after recalling the main properties of TRE and showing its connection
with QJSD, we present the key result of our paper, that is, the upper bound to the TRE revivals
in terms of the system-environment correlations and environmental-state changes; the proof of the bound
is provided in Appendix \ref{app:app}.
In Sec.\ref{sec:ex}, we apply our general analysis to two cases study and finally the conclusions
and outlooks of our work are given in Sec.\ref{sec:con}.

\section{Information backflow and non-Markovianity}\label{sec:ibm} 
Let us start recalling the basic properties of the TD useful for our analysis. The
TD is defined as $\mathsf{D}(\varrho,\sigma)= \sfrac 12\text{Tr}|\varrho-\sigma|$,
and provides a natural distance on the space of statistical
operators \cite{Holevo2001}.  Its crucial feature allowing to define and identify
memory effects is its contractivity under the action of a (completely) positive
trace preserving ((C)PT) map $\Phi$; namely TD obeys the so-called data
processing inequality
\begin{equation}\label{shrinkTr}
\mathsf{D}(\Phi[\varrho], \Phi[\sigma])\leqslant \mathsf{D}(\varrho,\sigma) 
\end{equation}
for any pair of states $\varrho,\sigma$~\footnote{Mostly, the data processing inequality is stated for complete positive maps. However, for trace distance and quantum relative entropy the positivity of the map is sufficient \cite{khatri2020principles}.}. Importantly, this property brings with itself invariance
under unitary maps and with respect to
the tensor product, i.e., $\mathsf{D}(\varrho,\sigma)=
\mathsf{D}(\varrho\otimes \tau ,\sigma\otimes \tau)$ for any state $\tau$,
as can be readily seen using CPT of both the partial
trace and the map $\varrho \mapsto \varrho\otimes \tau$. In
particular, TD is a proper quantum $f$-divergence, which allows us to
use it to quantify the distinguishability between quantum states
\cite{Fumio2017a}.
Finally, we consider two further important properties of the TD, which
are an immediate consequence of its being a distance in the
mathematical sense. The first is the validity of the triangle
inequality, which can be expressed as
\begin{align}
  \mathsf{D}(\varrho,\sigma)-\mathsf{D}(\varrho,\tau) &\leqslant \mathsf{D}(\sigma, \tau)
 \label{triangleTr1} \\
\mathsf{D}(\varrho,\sigma)-\mathsf{D}(\eta,\sigma) &\leqslant \mathsf{D}(\varrho,\eta),\label{triangleTr2}
\end{align}
for arbitrary states $\varrho$, $\sigma$, $\tau$ and $\eta$. The
second is positivity and boundedness according to
$0\leqslant\mathsf{D}(\varrho,\sigma)\leqslant 1$, with the value 0
iff $\varrho=\sigma$, and 1 iff their supports are orthogonal.
\begin{figure}[]
  \includegraphics[width=.45\textwidth]{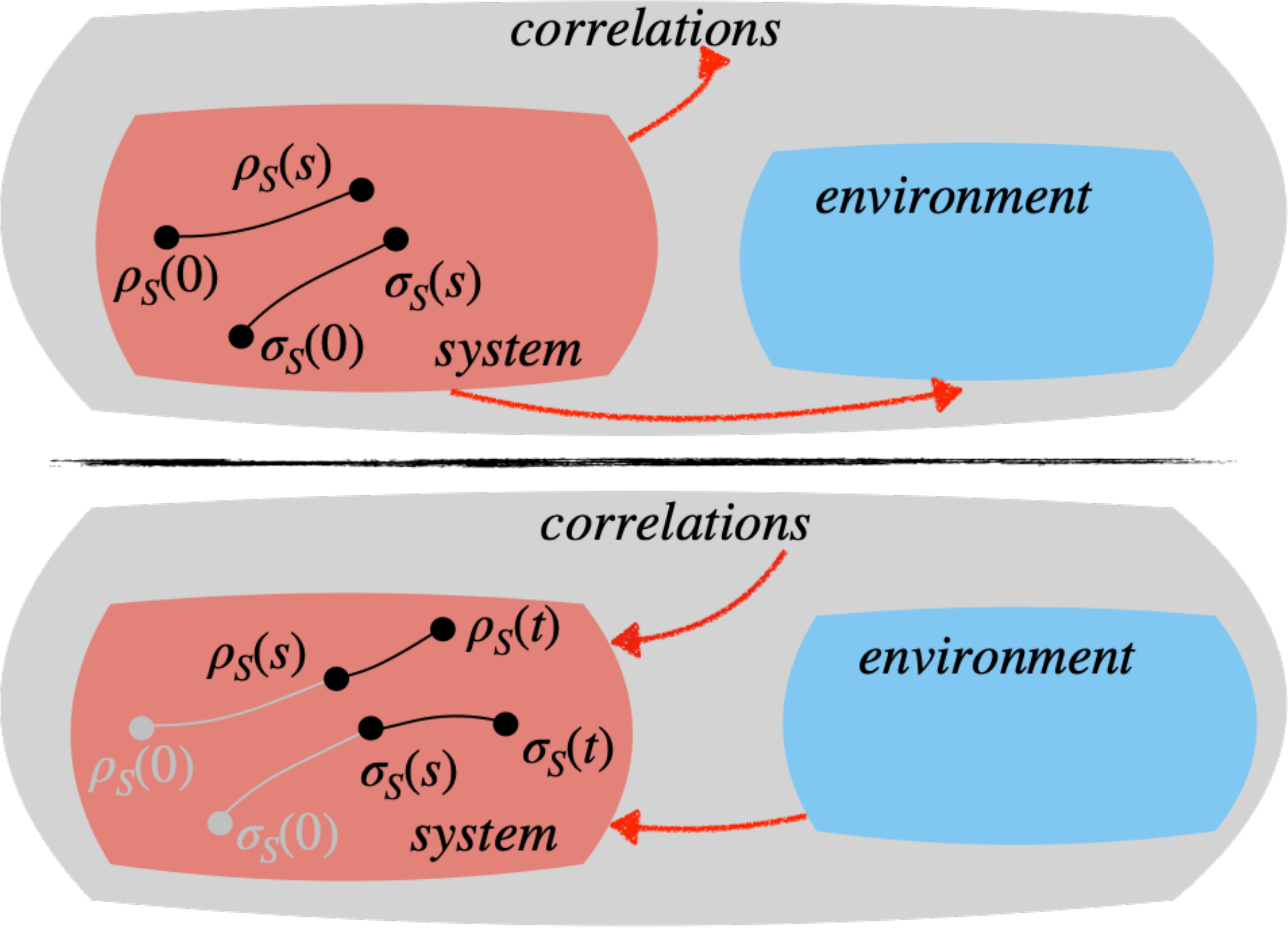}
  \caption{The concept of information backflow: initially the reduced states
    $\varrho_{\scriptscriptstyle S}$, $\sigma_{\scriptscriptstyle S}$
    approach each other since the information is flowing outside of
    the reduced system (top), while information backflow leads the states
    to diverge from each other (bottom).}
\label{Fig1}
\end{figure}
Thanks to these properties one can derive an upper bound
for the difference of the TD between the open system states
$\varrho_{\scriptscriptstyle S}$, $\sigma_{\scriptscriptstyle S}$, at
different times $t, s$: $t\geqslant s$. These states result
from two distinct initial conditions given by
factorized states with the same environmental marginal, $\varrho(0)=\varrho_{\scriptscriptstyle S}(0)\otimes \varrho_{\scriptscriptstyle E}(0)$, $\sigma(0)=\sigma_{\scriptscriptstyle S}(0)\otimes \sigma_{\scriptscriptstyle E}(0)=\sigma_{\scriptscriptstyle S}(0)\otimes \varrho_{\scriptscriptstyle E}(0)$, so as to ensure the existence of a reduced dynamics
\cite{Laine2010b,Amato2018a,Campbell_2019}. The bound reads:
\begin{align}\label{boundTr}
\!\!\!\!
  \mathsf{D}(\varrho_{\scriptscriptstyle S}(t),\sigma_{\scriptscriptstyle S}(t))-\mathsf{D}(\varrho_{\scriptscriptstyle S}(s),\sigma_{\scriptscriptstyle S}(s))\leqslant
  \mathsf{D}(\varrho_{\scriptscriptstyle
  E}(s),\sigma_{\scriptscriptstyle E}(s))+
  \nonumber
 \\ 
\mathsf{D}(\varrho_{}(s),\varrho_{\scriptscriptstyle S}(s)\otimes \varrho_{\scriptscriptstyle E}(s))+ \mathsf{D}(\sigma_{}(s),\sigma_{\scriptscriptstyle S}(s)\otimes \sigma_{\scriptscriptstyle E}(s)).
\end{align}
The interpretation of the above inequality is central to our analysis
and relies on the TD as a quantifier of distinguishability among
quantum states.  The terms at the right hand side of
Eq.~\eqref{boundTr} quantify the information about the initial state
of the open system, which we call local information, that is outside
the open system at time $s$, i.e., that can be accessed only via
measurements involving environmental degrees of freedom. Such
information can be encoded in the correlations between the open system
and the environment or in the environmental state. Indeed, although
  the environmental states are initially the same, they will generally
  differ at later times, due to the initial difference in the reduced
  states $\varrho_{\scriptscriptstyle S}(0)$ and
  $\sigma_{\scriptscriptstyle S}(0)$. On the other hand, the left
hand side of Eq.~\eqref{boundTr} concerns the accessibility of the
local information via measurements on the open system itself.  A positive value
of the difference means that the local information at time $t$ is greater than the one at time $s$, denoting a
backflow of information towards the open system. As the global system is closed, this
additional information must have an environmental origin, i.e., it was
previously contained in the environmental state or in the correlations
between system and environment, which is precisely what is shown by
Eq.~\eqref{boundTr}, see also Fig.~\ref{Fig1}.

We stress that, as should be clear from the previous analysis, any
distance contractive under CPT maps, e.g. the Bures distance considered
in \cite{Vasile2011a,Campbell_2019}, would lead to an inequality
analogous to the one in Eq.~\eqref{boundTr} and would then allow for
the same physical interpretation. However, till now no
entropic quantifier was used for this purpose, as no entropic distance
measure was known, though it was conjectured long ago that
the square root of the QJSD is actually a metric
\cite{PhysRevA.77.052311,Briet2009a}. This conjecture was recently
proven independently by two authors \cite{virosztek2019metric,sra2019metrics} and, accordingly, also the square root of the QJSD
is a suitable quantifier of the information backflow.  Here, we show
that there is a whole class of entropic quantities  that are in
general not distance measures but also allow for the interpretation
above; such a class includes QJSD as a special
case.  Thus, we substantiate the fact that the actual physical mechanism behind
the occurrence of memory effects in quantum dynamics is the
establishment of correlations or changes in the environmental states
upper bounding the revivals in local distinguishability.

\section{Telescopic relative entropy}\label{sec:tre}
Relative entropy is a fundamental quantity in statistical mechanics and information theory, both at classical and quantum level \cite{Wehrl1978,Schumacher2000a,Vedral2002a}. It also plays a distinguished role in quantum thermodynamics and its foundations, especially in analyzing the formulation of the second law of thermodynamics in the quantum regime \cite{SAGAWA_2012,Esposito2010a,Ptaszynski2019a,Floerchinger2020a}. 
The expression of the QRE first introduced by
Umegaki \cite{Umegaki1962a} reads $S(\varrho,\sigma)=\text{Tr}(\varrho\log\varrho-\varrho\log\sigma)$.
As well known, however, the QRE, while being the most relevant quantum $f$-divergence distinguishing quantum states \cite{Fumio2017a}, is not bounded and can diverge also in finite dimension. To cure this difficulty, regularised versions have been proposed \cite{RegEntropy,PhysRevA.72.052310,Audenaert1}. In particular we will show that the TRE introduced in \cite{Audenaert1} obeys an analog of the key inequalities Eq.~\eqref{shrinkTr}--\eqref{boundTr}. For a special choice of the telescopic parameter introduced below, the symmetrised version of TRE reduces to the QJSD, and a simplified upper bound for its square root follows.
With this, also for this class of entropic quantifiers an interpretation of the distinguishability revivals in terms of information backflow is given, thus fully justifying its use for the description of memory effects in an open quantum system dynamics.

The TRE is defined as
\begin{align}\label{Telesc}
\mathsf{S}_\mu(\varrho,\sigma)={\log(1/\mu)}^{-1} S(\varrho,\mu\varrho+(1-\mu)\sigma)
\end{align}
and is actually independent of the logarithm basis used in the
definition.
The telescopic parameter $\mu\in(0,1)$ gives the amount of mixing
between the two states $\varrho$ and $\sigma$, telling how much one
state is brought closer to the other moving along the joining line in
the convex set of states.
For the special choice of telescopic parameter $\mu={1}/{2}$ the symmetrised TRE $\mathsf{J}(\varrho,\sigma)={1}/{2}( S_{{1}/{2}}(\varrho,\sigma)+ S_{{1}/{2}}(\sigma,\varrho))$ equals the QJSD \cite{PhysRevA.72.052310}:
\begin{equation}\label{Jensen}
  \mathsf{J}(
  \varrho,\sigma)=\frac{1}{2}\left( S \left (\varrho,\frac{\varrho+\sigma}{2}\right)+ S \left(\sigma,\frac{\varrho+\sigma}{2}\right) \right) .
\end{equation}
The main property of the TRE, which distinguishes it from the standard QRE, is its boundedness. In particular, the prefactor is
chosen so that $0\leqslant \mathsf{S}_\mu(\varrho,\sigma)\leqslant  1$,
assuming the extreme values if and only if the states are identical or have orthogonal support \cite{Audenaert1}. What is more, TRE inherits from the QRE the joint convexity and the contractivity under (C)PT maps \cite{Hermes_2017}
\begin{align}\label{shrinkTRE}
\mathsf{S}_\mu(\Phi[\varrho], \Phi[\sigma])\leqslant \mathsf{S}_\mu(\varrho,\sigma),
\end{align}
thus being invariant under unitary transformations and tensor product $\mathsf{S}_\mu(\varrho,\sigma)= \mathsf{S}_\mu(\varrho\otimes \tau
  ,\sigma\otimes \tau)$.
Neither QRE nor TRE are distances as they do not satisfy the triangular inequality and are not symmetric in their arguments. However, it can be shown that TRE obeys the following inequalities \cite{Audenaert2}, similar in spirit to Eq.~\eqref{triangleTr1} and Eq.~\eqref{triangleTr2}:
\begin{align}\label{triangleTRE}
\mathsf{S}_\mu(\varrho,\sigma)-\mathsf{S}_\mu (\varrho,\tau)&\leqslant 1-\mathsf{S}_\mu(1,\mathsf{D}(\sigma,\tau)),\\
\mathsf{S}_\mu(\varrho,\sigma)-\mathsf{S}_\mu (\eta,\sigma)&\leqslant\mathsf{D}(\varrho,\eta)-\mathsf{S}_\mu(\mathsf{D}(\varrho,\eta),1) \label{triangle1},
\end{align}
where we have generalised the definition of TRE to act on non-negative
scalars in the obvious way.  The TRE can be bounded from below and above by
functions of the TD
\begin{equation}\label{Pinsker}
{2(1-\mu)^2} {\log(1/\mu)}^{-1}\mathsf{D}^2(\varrho,\sigma)\leqslant\mathsf{S}_\mu(\varrho,\sigma)\leqslant \mathsf{D}(\varrho,\sigma),
\end{equation}
where the lower bound is a straightforward generalization of the Pinsker inequality for the QRE \cite{Pinsker1960a}, but the upper bound is only possible since the TRE is bounded.

Exploiting these properties, we derive, as detailed in Appendix \ref{app:app}, the following inequality for the change in TRE
\begin{eqnarray}\label{boundTRE}
 &{\mathsf{S}}(\varrho_{\scriptscriptstyle S}(t),\sigma_{\scriptscriptstyle S}(t))-{\mathsf{S}}(\varrho_{\scriptscriptstyle S}(s),\sigma_{\scriptscriptstyle S}(s))  
  \leqslant
\kappa
  \Big(\sqrt[4]{{\mathsf{S}}(\varrho_{\scriptscriptstyle
        E}(s),\sigma_{\scriptscriptstyle E}(s))}+
   \nonumber
  \\
  &\!\! \!\! \!\! \!\!
    \sqrt[\leftroot{-2}\uproot{2}4]{\mathsf{S}(\varrho_{}(s),\varrho_{\scriptscriptstyle
  S}(s)\otimes \varrho_{\scriptscriptstyle E}(s))}
  \!+\!
  \sqrt[\leftroot{-2}\uproot{2}4]{\mathsf{S}(\sigma_{}(s),\sigma_{\scriptscriptstyle S}(s)\otimes \sigma_{\scriptscriptstyle E}(s))} \Big)
\end{eqnarray}
with $\mathsf{S}$ the TRE with telescopic parameter
$\mu={\rm e}^{-\sfrac 32}$. While boundedness of the TRE allows to
introduce a well defined non-Markovianity measure as for the TD
\cite{Breuer2016a}, this bound permits a full-fledged interpretation
of TRE as a quantifier of information backflow. This is true even
if TRE is not a distance, i.e. neither the triangle equality nor the
symmetry property are crucial for such an attribution.  Note that a
similar inequality holds for any telescopic parameter
$\mu$, see Appendix \ref{app:app}.  The value $\mu={\rm e}^{-\sfrac 32}$ corresponds to
the optimal one since it minimizes the prefactor 
(here, $\kappa=({{4 {\rm e}^3}/{27}})^{1/4}\approx 1.31$), and will be taken
as reference value.
\begin{figure}[]%
\includegraphics[width=0.45\textwidth]{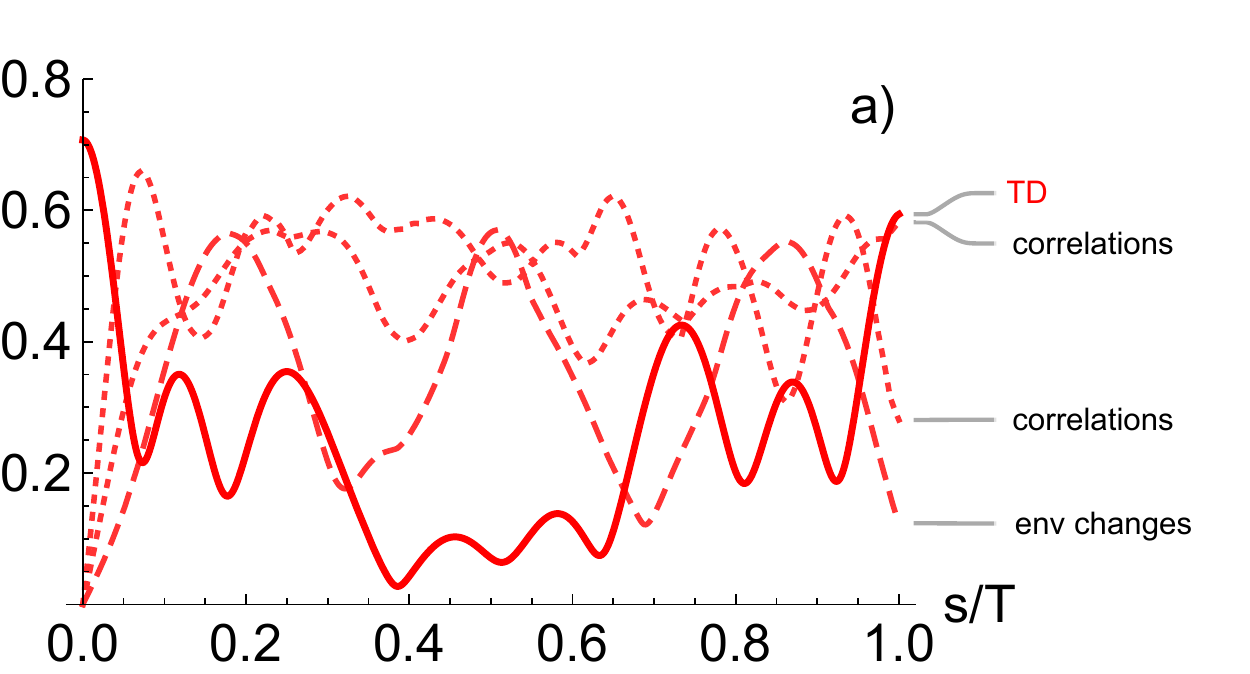}
\includegraphics[width=0.45\textwidth]{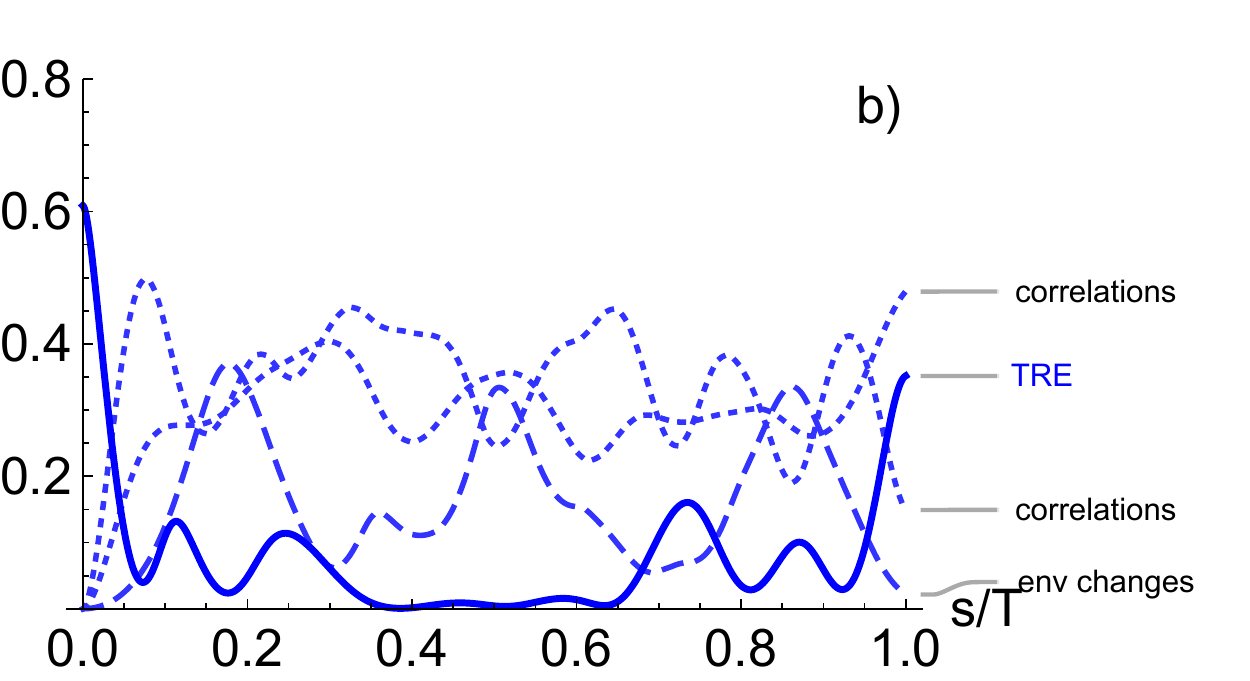}
\includegraphics[width=0.45\textwidth]{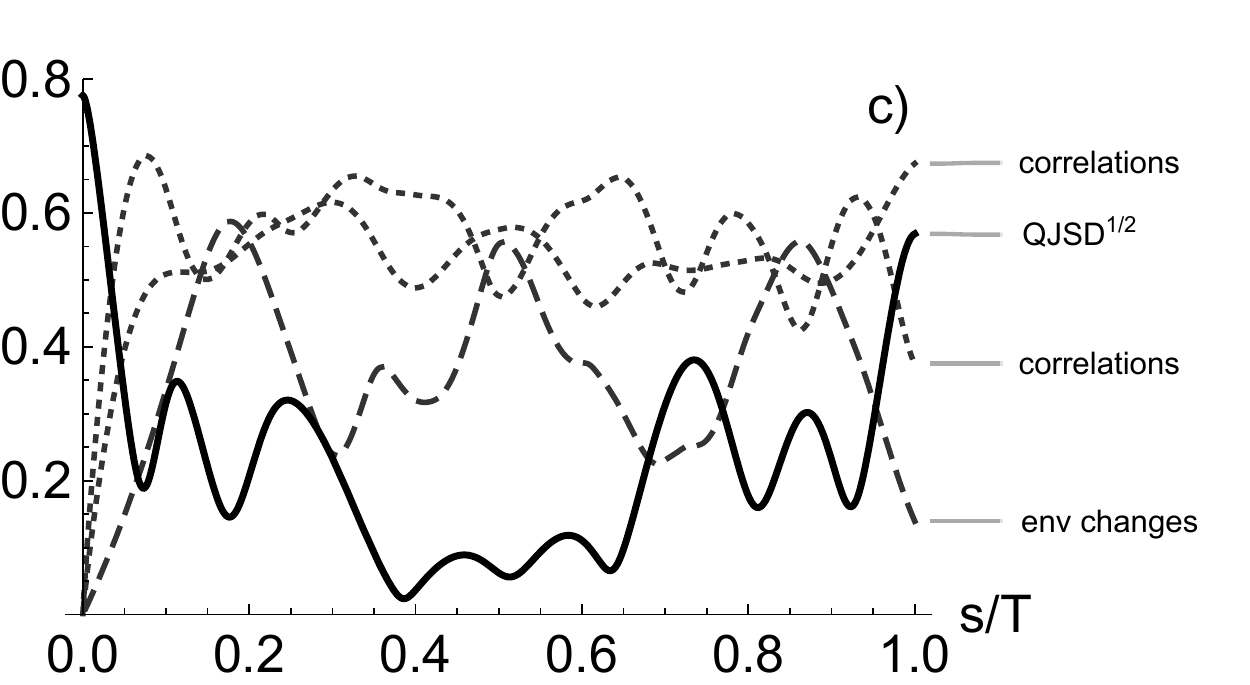}
\caption{%
  Different contributions to the bounds
  for the considered distinguishability quantifiers: $\mathbf{a)}$ the
  solid line is the TD as function of the rescaled time; the dashed
  line is the first contribution at the
  r.h.s. of Eq.~\eqref{boundTr} corresponding to changes in the
  environmental states, while the two dotted lines correspond to
  system-environment correlations; $\mathbf{b)}$ and $\mathbf{c)}$ provide the
  corresponding quantities relative to TRE and QJSD$^{\sfrac 12}$.
  The initial states of
  the qubit are given by the up state and a
  symmetric superposition of up and down state, while the environment
  starts in a thermal state with $\beta \omega_{\scriptscriptstyle E}=1$. We set $T=8.9$ in
  inverse units of the coupling strength $g$.}
\label{Fig2}
\end{figure}
For $\mu={1}/{2}$, the symmetrised TRE corresponds to the QJSD $\mathsf{J}(\varrho,\sigma)$, Eq.~\eqref{Jensen}, and it was recently shown, more than 10 years after the proof for the classical Jensen-Shannon divergence \cite{Endres2003a} and the conjecture for the quantum case \cite{PhysRevA.77.052311,Briet2009a}, that its square root is a proper distance \cite{virosztek2019metric, sra2019metrics}. With this and the contractivity under CPT maps one immediately has
\begin{eqnarray}\label{boundQJSD}
 &\sqrt[\leftroot{-2}\uproot{2}]{\mathsf{J}(\varrho_{\scriptscriptstyle
     S}(t),\sigma_{\scriptscriptstyle
     S}(t))}- \sqrt[\leftroot{-2}\uproot{2}]{\mathsf{J}(\varrho_{\scriptscriptstyle
     S}(s),\sigma_{\scriptscriptstyle S}(s))}\leqslant
 \sqrt[\leftroot{-2}\uproot{2}]{\mathsf{J}(\varrho_{\scriptscriptstyle
  E}(s),\sigma_{\scriptscriptstyle E}(s))}+
  \nonumber
 \\
  &\!\! \!\! \!\! \!\! \!\! 
    \sqrt[\leftroot{-2}\uproot{2}]{\mathsf{J}(\varrho_{}(s),\varrho_{\scriptscriptstyle S}(s)\otimes \varrho_{\scriptscriptstyle E}(s))}\!+ \sqrt[\leftroot{-2}\uproot{2}]{\mathsf{J}(\sigma_{}(s),\sigma_{\scriptscriptstyle S}(s)\otimes \sigma_{\scriptscriptstyle E}(s)))}.
\end{eqnarray}

\section{Examples}\label{sec:ex} 
\begin{figure}[]
 \hspace*{.8cm}
      \includegraphics[width=0.45\textwidth]{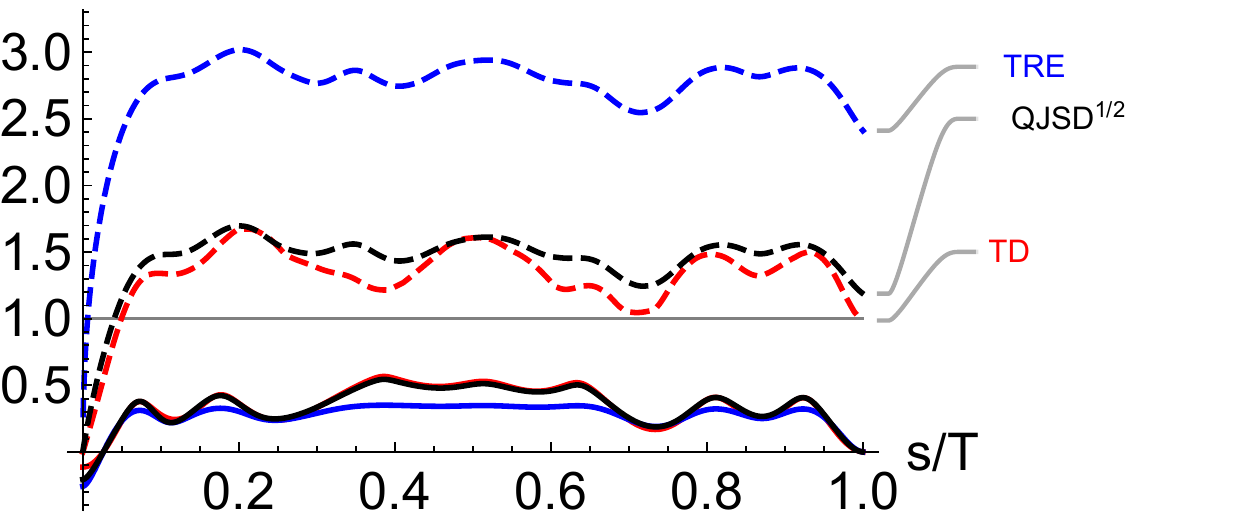}
    \caption{%
   Revivals of
   distance and entropic distinguishability quantifiers versus their
   bounds in terms of correlations and environmental changes for the
   Jaynes–Cummings model: TD (red), TRE (blue) and QJSD$^{\sfrac 12}$ (black).
   Solid and dashed lines correspond to l.h.s. and r.h.s. of
   Eqs.~\eqref{boundTr}, \eqref{boundTRE} and \eqref{boundQJSD}
   respectively.    The straight line at 1 corresponds to
   the maximal possible value of the revivals. The value of $t$ at the
   l.h.s. is set to $T=8.9$ in inverse units of the coupling strength
   $g$, corresponding to a local maximum of the 
   distinguishability as in
   Fig.~\ref{Fig2}. The very close
   behavior of TD and QJSD$^{\sfrac 12}$ clearly appears. All
   parameters are as in Fig.~\ref{Fig2}.}
 \label{Fig3}
\end{figure}

We now showcase our findings by examples.
Let us consider first the Jaynes–Cummings model, describing
the interaction of a qubit with a single bosonic field mode
\begin{align}
H=\omega_{\scriptscriptstyle S} \sigma_z \otimes \mathbb{I} + g ( \sigma_+\otimes b+ \sigma_- \otimes b^\dagger)+ \omega_{\scriptscriptstyle E} \mathbb{I}\otimes  b^\dagger b,
\end{align}
where we introduced the raising and lowering operators
$\sigma_{ \pm}=\sigma_x \pm i \sigma_y$ expressed in terms of the
Pauli matrices, while $b, b^\dagger$ are bosonic creation and
annihilation operators, respectively.  This model can be solved
exactly, see \cite{Puribook, PhysRevA.82.022110}, thus allowing for a
comparison of the TD and TRE quantities occurring in
Eqs.~\eqref{boundTr}, \eqref{boundTRE} and \eqref{boundQJSD}. In
Fig.~\ref{Fig2} we report in each panel the l.h.s. and the three
contributions at the r.h.s. of the bounds for the TD, the TRE and the
square root of the QJSD, which we denote as QJSD$^{\sfrac 12}$,
respectively.  We see that the qualitative behaviour of these
quantifiers of quantum-state distinguishability is similar, with
respect to both the information contained within the open system and
the one outside it, namely the system-environment correlations and
environmental states. The quantities referred to the
QJSD$^{\sfrac 12}$ in particular mimic very tightly the behaviour of
the corresponding TD quantities. The TRE is always smaller than the
corresponding TD, which is a general feature for all telescopic
parameters $\mu$, see Eq.~\eqref{Pinsker}. From this, however, one
cannot conclude that the terms appearing in Eq.~\eqref{boundTRE} are
always smaller than the corresponding ones in Eq.~\eqref{boundTr}, as
can be seen in Fig.~\ref{Fig3}. Actually, the upper bounds in terms of
the entropic quantities are almost always less tight for this model
than the corresponding TD one. However, as all three bounds are for
most of the time above one, their applicability for estimation of the
l.h.s., which is never larger than one, is rather
limited. Nonetheless, their existence guarantees the direct relation
to information backflow, as the revival of local distinguishability
unambiguously originates from establishment of correlations or changes
in the environmental states. They thus act as precursors of
non-Markovianity \cite{Campbell_2019} and the assessment of the
different contributions allows to infer which is the most relevant
physical mechanism behind the revivals, whose time dependence is
mirrored in the bound.

\begin{figure}[]
 \includegraphics[width=0.45\textwidth]{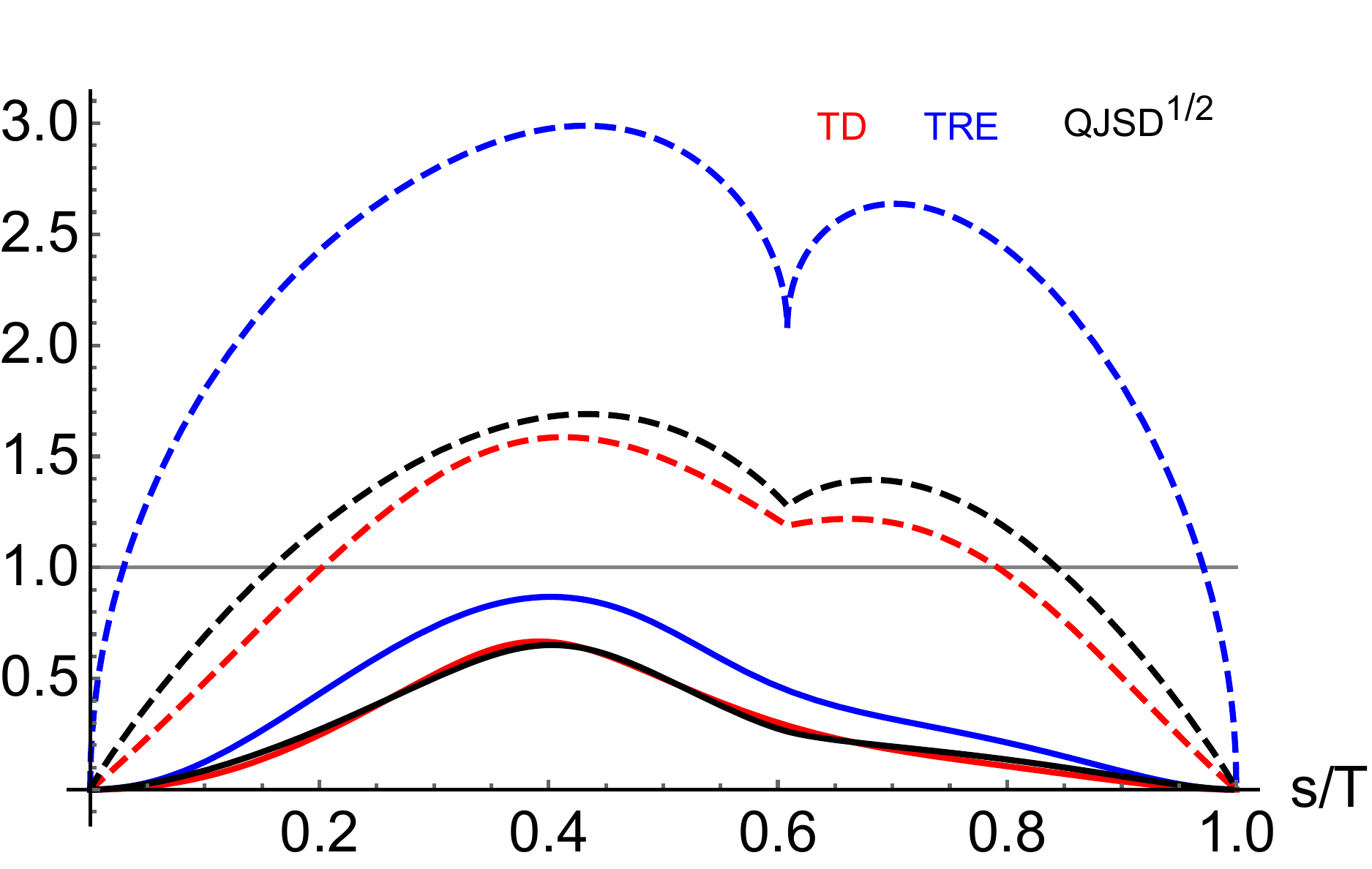}
 \caption{Revivals in distinguishability and
   their upper bounds, similarly to Fig.~\ref{Fig3} but for the case of
   the dissipative qubit model. Also in this case solid and dashed lines
   correspond to l.h.s. and r.h.s. of Eqs.~\eqref{boundTr},
   \eqref{boundTRE} and \eqref{boundQJSD}, for TD (red),
   TRE (blue) and QJSD$^{\sfrac 12}$
   (black) respectively.  The initial states are chosen to be pure orthogonal
   states in the $yz$ plane of the Bloch sphere. The reference time
   $t$ is here set to $T=\pi$ in inverse units of the coupling
   $g$. The environmental state is also taken to be pure in the $xz$
   plane.}
  \label{Fig4}
\end{figure}

\begin{figure}[]
 \includegraphics[width=0.45\textwidth]{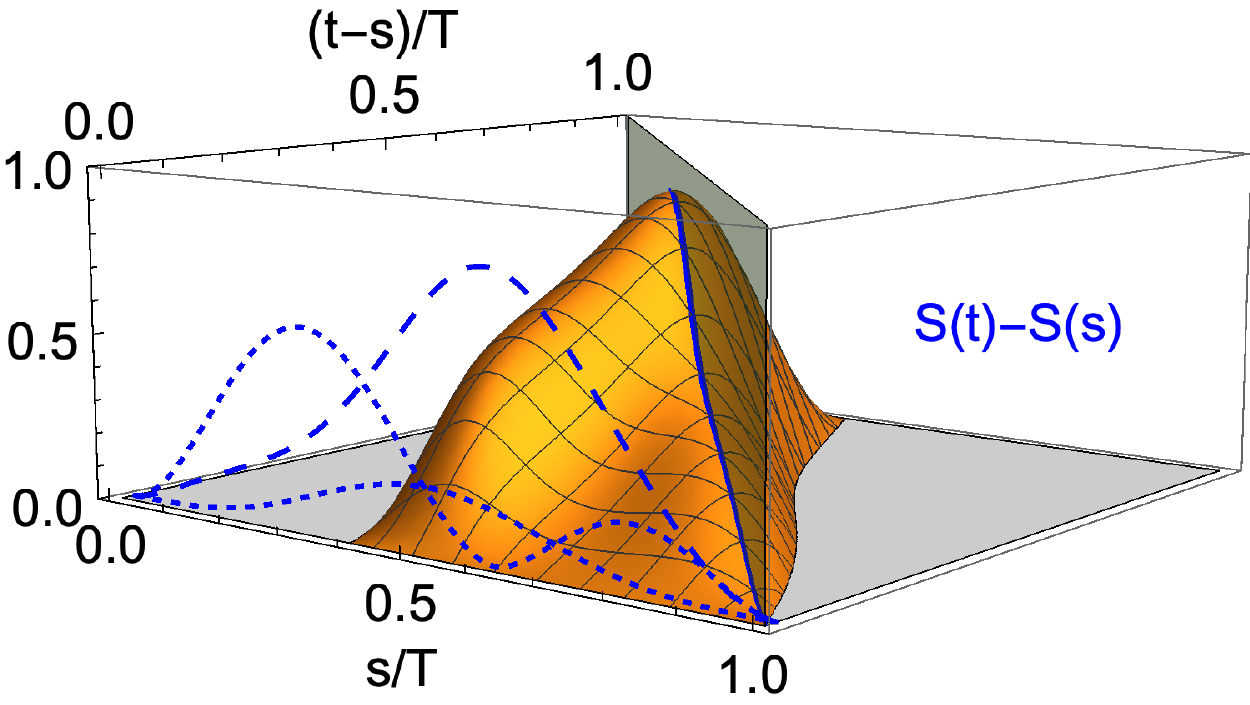}\\
\includegraphics[width=0.45\textwidth]{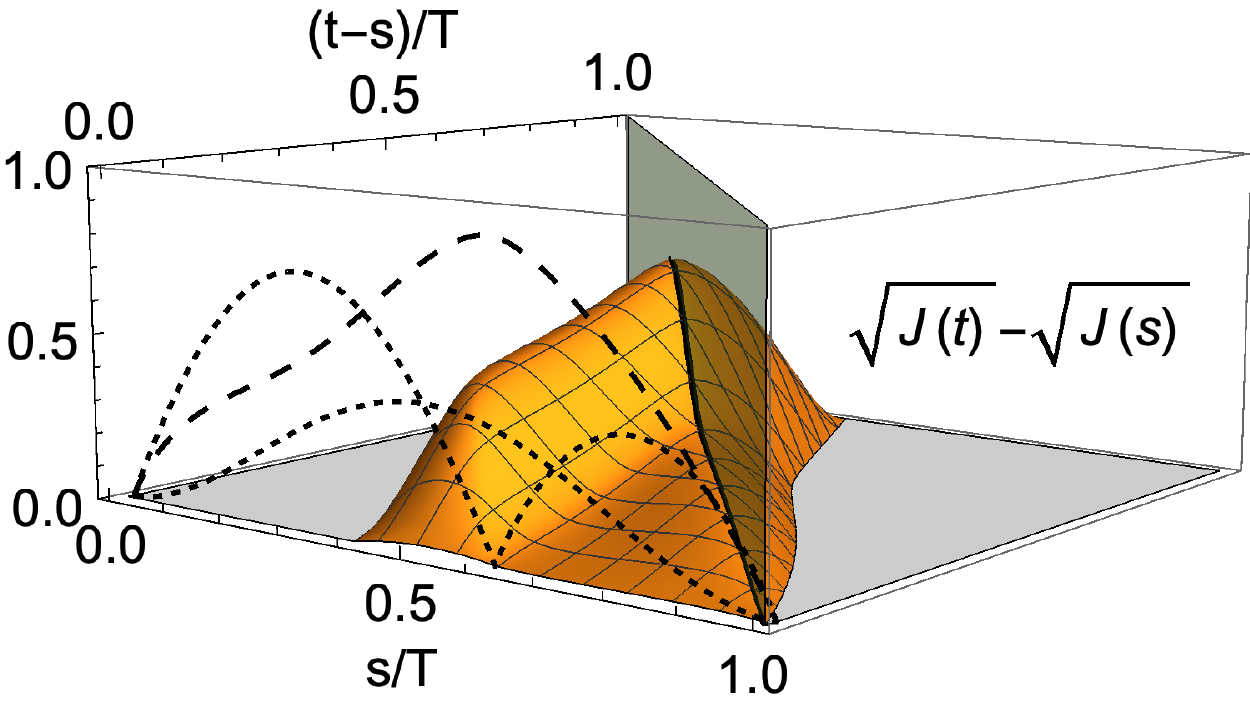}
\caption{Comparison of TRE (top)
  and QJSD$^{\sfrac 12}$ (bottom) for the two qubit dissipative
  dynamics with
  initial states as in Fig.~\ref{Fig2}. The dashed lines in the $s=t$
  plane correspond to the difference in the environmental states, the
  dotted lines to the correlations. Solid lines in the section
  correspond to TRE and QJSD$^{\sfrac 12}$ as in
  Fig.~\ref{Fig4}. The reference time $T$ is set to $\pi$ in
  inverse units of the coupling $g$.}
\label{Fig5}
\end{figure}

For better comparison we also consider a simpler model, and set the
environment to be a qubit in resonance with the reduced one, see
\cite{Tang2014}, with interaction term $g ( \sigma_+\otimes \sigma_- + \sigma_- \otimes \sigma_+)$.
Fig.~\ref{Fig4} shows that the behaviour of TD and entropic quantities
is again qualitatively similar, especially the QJSD$^{\sfrac 12}$
follows quite closely the corresponding TD quantities. What is more,
the TD bound is once again tighter than the corresponding entropic
bounds. However, due to the occurrence of the roots in
Eq.~\eqref{boundTRE} and Eq.~\eqref{boundQJSD} one clearly notes that
the latter are more sensitive to the changes in the system-environment
correlations for times around $t=0.6 T$, when one of the two global
states factorizes, as shown in Fig.~\ref{Fig5}, where we compare the
behaviour of the TRE and QJSD$^{\sfrac 12}$ quantities.

\section{Conclusions}\label{sec:con} 
We have shown that entropic quantities can be used to consistently
define the exchange of information between an open
quantum system and its environment. By focusing on a class of
regularised versions of the quantum relative entropy, named telescopic
relative entropy, we derived an upper bound to the variation of the
reduced state distinguishability in terms of the information lying
outside the open system, encoded in the system-environment
correlations and the environmental states. Besides strengthening the
interpretation of non-Markovianity as backflow of information, our
results also clarify which are the key mathematical properties behind
this picture.  Furthermore, we showed that a special case of the
telescopic relative entropy can be connected to the quantum
Jensen-Shannon divergence. The square root of the latter yields a proper metric on the
set of quantum states and it can reproduce both qualitatively and
quantitatively the behavior of the trace distance.  These features
have been highlighted by means of examples. 

In future investigations it will be important to
understand to what extent the use of entropic quantities to
characterize non-Markovian open system dynamics can be further
justified, developing a measure of non-Markovianity by the
detection of the pair of states maximizing the backflow of information
\cite{BLP,Wissmann2012,Liu2014}, or connecting the revivals of
distinguishability with the presence of initial correlations \cite{Dajka2011}.

\begin{acknowledgments}
NM acknowledges funding by the Alexander von Humboldt Foundation in form of a Feodor-Lynen
Fellowship. All authors acknowledge support
from the UniMi Transition Grant H2020.
\end{acknowledgments}

\onecolumngrid

\appendix

\section{Proof of Eq.~\eqref{boundTRE}}\label{app:app}
We prove validity of the inequality Eq.~\eqref{boundTRE}, relying on the already introduced properties of the TRE.  This inequality provides an upper bound for the revivals of the TRE for the reduced states of the system at different times, showing that this upper bound is due to the establishment of correlations between system and environment, as well as changes in the environmental states. Importantly, in the absence of these modifications the bound is equal to zero. 

Fixed a pair of initial system states $\varrho_{\scriptscriptstyle S}(0)$ and $\sigma_{\scriptscriptstyle S}(0)$, we introduce the following notation for the quantity at the l.h.s. of Eq.~\eqref{boundTRE} \begin{equation}
  \label{eq:1}
I_\mu(t,s)  \equiv \mathsf{S}_\mu(\varrho_{\scriptscriptstyle S}(t),\sigma_{\scriptscriptstyle S}(t))-\mathsf{S}_\mu(\varrho_{\scriptscriptstyle S}(s),\sigma_{\scriptscriptstyle S}(s)).
\end{equation}
Exploiting CPT of the partial trace and invariance of TRE under unitaries, with the natural assumption of unitarity of the overall evolution we can write
\begin{align}\label{step1}
I_\mu(t,s)\leqslant \mathsf{S}_\mu(\varrho_{}(t),\sigma_{}(t) )-\mathsf{S}_\mu(\varrho_{\scriptscriptstyle S}(s),\sigma_{\scriptscriptstyle S}(s))
 =\mathsf{S}_\mu(\varrho_{}(s),\sigma_{}(s))-\mathsf{S}_\mu(\varrho_{\scriptscriptstyle S}(s),\sigma_{\scriptscriptstyle S}(s)).
\end{align}
Adding and subtracting the quantity $\mathsf{S}_\mu(\varrho_{\scriptscriptstyle S}(s)\otimes\varrho_{\scriptscriptstyle E}(s) ,\sigma_{}(s))$ at the r.h.s. and using invariance under tensor product we are left with
\begin{align}
    \label{eq:2bis}
  I_\mu(t,s) \leqslant |\mathsf{S}_\mu(\varrho_{}(s),\sigma_{}(s)) - \mathsf{S}_\mu(\varrho_{\scriptscriptstyle S}(s)\otimes\varrho_{\scriptscriptstyle E}(s) ,\sigma_{}(s))|+|\mathsf{S}_\mu(\varrho_{\scriptscriptstyle S}(s)\otimes\varrho_{\scriptscriptstyle E}(s) ,\sigma_{}(s)) -\mathsf{S}_\mu(\varrho_{\scriptscriptstyle S}(s)\otimes \varrho_{\scriptscriptstyle E}(s),\sigma_{\scriptscriptstyle S}(s)\otimes \varrho_{\scriptscriptstyle E}(s))|.
  \end{align}
To proceed further we rearrange the triangle-like
inequalities Eqs.~\eqref{triangleTRE} and \eqref{triangle1} according to
\begin{equation}
  \label{eq:3}
  \mathsf{S}_\mu(\varrho,\sigma)-\mathsf{S}_\mu(\varrho,\tau)\leqslant \frac{1}{\log(1/\mu)}\log\left( 1+\mathsf{D}(\sigma,\tau)\frac{1-\mu}{\mu} \right)
\end{equation}
and
\begin{equation}
  \label{eq:4}
  \mathsf{S}_\mu(\varrho,\sigma)-\mathsf{S}_\mu(\eta,\sigma)\leqslant  \frac{\mathsf{D}(\varrho,\eta)}{\log(1/\mu)}\log\left( 1+\frac{1}{\mathsf{D}(\varrho,\eta)}\frac{1-\mu}{\mu} \right),
\end{equation}
which provide more convenient starting points for upper bounding the different contributions. Indeed we can now
exploit  the following inequality, valid for non-negative  $x$ \cite{Topsoe}
\begin{equation}
  \label{topsoe}
\log(1+x)\leqslant \frac{x}{\sqrt{1+x}},
\end{equation}
entailing as special case
\begin{equation}
  \label{eq:3a}
  \log(1+x)\leqslant \sqrt{x}.
\end{equation}
We now exploit Eqs.~\eqref{eq:4} and \eqref{topsoe} 
for the first term at the r.h.s. of Eq.~\eqref{eq:2bis}
upon the identification $\sigma \rightarrow \sigma_{}(s)$, as well as Eqs.~\eqref{eq:3} and \eqref{eq:3a} 
for the second term at the r.h.s. of Eq.~\eqref{eq:2bis} to obtain
\begin{align}\label{step4}
I_\mu(t,s) \leqslant
 \sqrt{\frac{1-\mu}{\mu}}\frac{1}{\log(1/\mu)} \left(\sqrt{\mathsf{D}(\varrho_{}(s),\varrho_{\scriptscriptstyle S}(s)\otimes\varrho_{\scriptscriptstyle E}(s))} +\sqrt{\mathsf{D}(\sigma_{}(s),\sigma_{\scriptscriptstyle S}(s)\otimes \rho_{\scriptscriptstyle E}(s))}\right).
\end{align}
We further use
the triangle inequality and the invariance of the TD with respect to tensor product, together with the inequality  $\sqrt{x+y} \leqslant \sqrt{x}+\sqrt{y}$ valid for non-negative $x$ and $y$, thus finally coming to
\begin{align}\label{step44}
I_\mu(t,s) \leqslant
 \sqrt{\frac{1-\mu}{\mu}}\frac{1}{\log(1/\mu)} \left(\sqrt{\mathsf{D}(\varrho_{}(s),\varrho_{\scriptscriptstyle S}(s)\otimes\varrho_{\scriptscriptstyle E}(s))} +\sqrt{\mathsf{D}(\sigma_{}(s),\sigma_{\scriptscriptstyle S}(s)\otimes \sigma_{\scriptscriptstyle E}(s))}+\sqrt{\mathsf{D}(\sigma_{\scriptscriptstyle E}(s),\rho_{\scriptscriptstyle E}(s))}\right).
\end{align}
As a last step we use the generalized Pinsker inequality Eq.~\eqref{Pinsker} and obtain the upper bound
\begin{align}\label{step5}
I_\mu(t,s) \leqslant 
\frac{1}{\sqrt[4]{2\mu^2\log^3(1/\mu)}}   \left( \sqrt[4]{\mathsf{S}_\mu(\varrho_{}(s),\varrho_{\scriptscriptstyle S}(s)\otimes\varrho_{\scriptscriptstyle E}(s))} 
 + \sqrt[4]{\mathsf{S}_\mu(\sigma_{}(s),\sigma_{\scriptscriptstyle S}(s)\otimes \sigma_{\scriptscriptstyle E}(s))}+\sqrt[4]{\mathsf{S}_\mu(\sigma_{\scriptscriptstyle E}(s),\varrho_{\scriptscriptstyle E}(s))} \right).
\end{align}
Note that the prefactor ${1}/{\sqrt[4]{2\mu^2\log^3(1/\mu)}}$ as a
function of $\mu$ has a global minimum $({{4 {\rm
      e}^3}/{27}})^{1/4}\approx 1.31$ at $\mu=({1}/{\rm
  e})^{\frac{3}{2}}$. This is the choice of telescopic parameter
considered in  Eq.~\eqref{boundTRE}, which is now proven.

Let us mention the fact that different upper bound of the triangle-like
inequalities Eqs.~\eqref{eq:3} and \eqref{eq:4} can be considered, leading to another bound with different functional dependency on the difference in environmental states and correlations.The bound reads
\begin{align}\label{boundTRE2}
I_\mu(t,s) 
\leqslant \frac{1}{\sqrt{2\mu^2\log(1/\mu)}}
 \left( \sqrt[4]{\mathsf{S}_\mu(\varrho_{}(s),\varrho_{\scriptscriptstyle S}(s)\otimes\varrho_{\scriptscriptstyle E}(s))} + \sqrt[4]{\mathsf{S}_\mu(\sigma_{}(s),\sigma_{\scriptscriptstyle S}(s)\otimes \sigma_{\scriptscriptstyle E}(s))}+\sqrt{\mathsf{S}_\mu(\sigma_{\scriptscriptstyle E}(s),\varrho_{\scriptscriptstyle E}(s))} \right),
\end{align}
where the prefactor is obtained by starting from the looser bound $\log(1+x)\leqslant x$.
The prefactor ${1}/\sqrt{2\mu^2\log(1/\mu)}$ has its minimum
$\sqrt{\rm e}$ at $\mu={1}/{\sqrt{\rm e}}$. Note that $\sqrt{\rm e}>({{4 {\rm e}^3}/{27}})^{1/4}$, accordingly {in the most common situations}, where during the greatest part of the dynamics of the open quantum system the creation of correlations plays the dominant role rather than the change in the environmental state, the bound given by Eq.~\eqref{boundTRE} is the tighter one.

\end{document}